\documentclass[aps,prl,twocolumn,groupedaddress,showpacs,showkeys]{revtex4}

\usepackage{graphicx}
\begin{document}

\newcommand{\be}{\begin{equation}}
\newcommand{\ee}{\end{equation}}

%Title of paper
\title{The thermal Casimir effect  for rough metallic plates}

\author{ Giuseppe Bimonte}
\email[Bimonte@na.infn.it]
%\thanks{}
%\altaffiliation{}
\affiliation{Dipartimento di Scienze Fisiche Universit\`{a} di
Napoli Federico II Complesso Universitario MSA, Via Cintia
I-80126 Napoli Italy and INFN Sezione di Napoli, ITALY\\
}

\date{\today}

\begin{abstract}

We   propose a new  theory of thermal  Casimir effect, holding for
the experimentally important case of metallic surfaces with a
roughness having a spatial scale smaller than the skin depth. The
theory is based on a simple phenomenological model for a rough
conductor, that explicitly takes account of the fact that ohmic
conduction in the immediate vicinity of the surface of a conductor
is much impeded by surface roughness, if the amplitude of
roughness is smaller than the skin depth. As a result of the new
model, we find that surface roughness strongly influences the
magnitude of the thermal correction to the Casimir force,
independently of the plates separation. Our model, while
consistent with recent accurate measurements of the Casimir force
in the submicron range, leads to a new prediction for the not yet
observed thermal correction to the Casimir force at large plates
separation.    Besides the thermal Casimir problem, our model is
relevant for the correct theoretical interpretation of current
experiments probing other proximity effects between conductors,
like radiative heat transfer and quantum friction.

% insert abstract here
\end{abstract}

% insert suggested PACS numbers in braces on next line
% insert suggested keywords - APS authors don't need to do this

%\maketitle must follow title, authors, abstract, \pacs, and \keywords
\maketitle
% body of paper here - Use proper section commands
% References should be done using the \cite, \ref, and \label commands

The Casimir effect \cite{bordag} provides physicists with the rare
opportunity to investigate in a laboratory the physics of the
quantum vacuum \cite{milonni}, a subject that  has recently
attracted much interest for a number of reasons. One one hand, the
possibility exists that the recently discovered accelerated
expansion of Universe might be explained by a cosmological term
originating from quantum vacuum effects \cite{peebles}. Another
reason of interest comes from current experiments on non-newtonian
forces at the sub-micron scale \cite{fisch}. As its is well known,
Casimir forces are the dominant forces at this scale, and
therefore determining them very accurately is indispensable in
order to obtain experimental bounds on possible non-Newtonian
forces \cite{decca}. Finally, we would like to mention that the
Casimir effect, and more in general van der Waals forces
\cite{parse}, are now finding new   applications in nanotechnology
\cite{capasso}.

In the field of Casimir physics, an important and much
controversial problem is that of determining how the Casimir force
between two metallic bodies is modified by the temperature of the
environment \cite{bordag}. Apart from its intrinsic interest as a
problem in theory of quantum electromagnetic (e.m.) fluctuations,
addressing this problem is important because many  experiments  on
non-newtonian forces at the submicron scale use metallic surfaces
at room temperature \cite{decca}. We remark that, while the
Casimir pressure has now been measured accurately
\cite{lamor,decca}, the thermal contribution to the Casimir effect
has not yet been detected as we write, and indeed there are at
present several ongoing and planned experiments, to measure it
\cite{brown}. Surprisingly, the theory of the thermal Casimir
effect for metallic bodies turned out to be rather complicated,
and at present there exist several conflicting approaches that
give widely different predictions for the magnitude of the effect
\cite{bordag}.

In this Letter we present a new theory of the thermal Casimir
effect  for conductors. The distinctive new feature of the theory
consists in the central role played in our approach by the surface
roughness of the conductor. The presence of roughness is indeed a
general feature of current Casimir experiments, which use metallic
surfaces presenting surface roughness with a characteristic
amplitude $h$ of a few nanometers \cite{bordag,decca}. Until now,
it has always been thought that for separations $a$ between the
surfaces much larger than the roughness amplitude $h$, surface
roughness implies only a small correction of order $(h/a)^2$
\cite{bordag}. We  argue that this conclusion, while correct for
insulators, is   not warranted for metallic surfaces at finite
temperature, if the roughness $h$ is smaller than the skin depth
$\delta$, because then account must be taken of the fact that
surface irregularities impede ohmic conduction in the immediate
vicinity of the surface, a fact noted already by Pippard
\cite{pippard} long ago. Using a simple phenomenological model to
account for this effect, we then find that {\it independently of
the separation $a$}, the magnitude of the thermal correction to
the Casimir force strongly depends on the surface roughness. While
consistent with a recent  precise measurement \cite{decca} of the
Casimir pressure in the submicron range, our theory leads to a new
prediction for the magnitude of the not yet measured thermal
correction to the Casimir force at large separations. We note that
the application range of our model  extends to other proximity
effects originating from e.m. fluctutations, like radiative heat
transfer \cite{polder} and quantum friction between closely space
metallic surfaces \cite{pendry}, that are currently under intense
investigation, but we shall not tough upon these issue in this
Letter.

All current approaches to the thermal Casimir effect are based on
the theory of the Casimir effect developed long ago by Lifshitz
\cite{lifs}, on the basis of Rytov's general theory of e.m.
fluctuations \cite{rytov}. Lifshitz-Rytov theory works very well
in the case of media that can be characterized by electric and
magnetic permittivity tensors, $\epsilon_{ij}(\omega)$ and
$\mu_{ij}(\omega)$ that depend only on the frequency, but are
independent of the wave vector ${\bf k}$, i.e. for media that
exhibit only time-dispersion but not space-dispersion. According
to this theory, the Casimir pressure $P$ between two identical
plane-parallel (possibly stratified) slabs at temperature $T$,
separated by an empty gap of width $a$, is: \be P=- \frac{k_B T}{2
\pi^2} \sum_{l \ge 0}{\,'} \int_0^{\infty}\!\! d^2{\bf k_{\perp}}
q_l\!\!\!\! \sum_{\alpha={\rm TE,TM}} \left(\frac{e^{2 a
q_l}}{r_{\alpha}^2(i \xi_l,{\bf k_{\perp}})}-1
\right)^{-1},\label{lifs} \ee where the prime over the $l$-sum
means that the $l=0$ term has to taken with a weight one half,
${\bf k_{\perp}}$ denotes the projection of the wave-vector onto
the plane of the plates and $q_l =\sqrt{k_{\perp}^2+\xi_l^2}$,
where $\xi_l= 2 \pi k_B T l/\hbar$. The quantities $r_{\alpha}(i
\xi_l,{\bf k_{\perp}})$ denote the reflection coefficients of the
slabs for $\alpha$-polarization (for simplicity, we do not
consider the possibility of a non-diagonal reflection matrix),
evaluated at complex frequencies $i \xi_l$. For the case of
single-layer homogeneous slabs, with permittivity
$\epsilon(\omega)$, $r_{\alpha}$ are the familiar Fresnel
reflection coefficients. The above formula for the Casimir
pressure is regularly used also in the case of metals, that can be
characterized by a (complex) conductivity $\sigma(\omega)$, in
which case one has for the permittivity the expression
$\epsilon(\omega)=1+ 4 \pi i\,\sigma(\omega)/\omega$. The
difficulties found recently in the study of the thermal Casimir
effect for metals originate from the fact that in order to
evaluate the $l=0$ term of Eq. (\ref{lifs}) one needs extrapolate
the reflection coefficients $r_{\alpha}$ to zero frequency.
Different prescriptions for doing this have been proposed, leading
to largely different predictions for the thermal correction to the
pressure, depending on the resulting magnitude of the $l=0$ term.
For example, the approach of Ref.\cite{sernelius} adopts the Drude
model \be \epsilon_D(\omega)=1-\Omega_P^2/[\omega(\omega+i
\gamma)]\;,\label{drude}\ee where $\Omega_P$ is the plasma
frequency and $\gamma$ the relaxation frequency accounting for
ohmic conductivity. When this model is used, one finds that the
$l=0$ term for TE polarization gives zero contribution, and the
predicted thermal correction is much larger than what one obtains
for the ideal case of perfect reflectors (corresponding to taking
$r_{\alpha}^2=1$ in Eq. (\ref{lifs})). Recently, it has been shown
that this model is inconsistent with the experiment \cite{decca}
and moreover it has been noted that, in the case of perfect
lattices, the Drude model leads to thermodynamical inconsistencies
at low temperature \cite{bezerra}. On the contrary, the approach
based on the plasma model \be
\epsilon_P(\omega)=1-\Omega_P^2/\omega^2\;,\ee (augmented by a
six-oscillator contribution accounting for interband transitions),
has been shown to be consistent with the experiment \cite{decca},
and it is immune of thermodynamic inconsistencies.

All the above is for perfectly plane-parallel plates while, as we
said earlier, real experiments involve rough surfaces, with
amplitude $h$ typically in the range of a few nanometers.  In the
existing literature, the effect of the slabs roughness has always
been regarded as a {\it geometric} problem, involving as the only
relevant parameter the geometric quantity $h/a$. For $h \ll a$,
the problem has been studied by perturbative means, for example
the so-called Proximity Force Approximation (PFA) which
corresponds to taking a suitable average of the Lifshitz formula
result, across the surface of the plates \cite{bordag}. The result
of the PFA is that for $h \ll a$, roughness gives only a small
fractional correction of order $(h/a)^2$ to the Casimir force
between two perfectly flat plates  (see for example Ref.
\cite{decca}). Below we show that such a conclusion is not correct
however, if the roughness profile varies rapidly along the surface
of the conductor,  over distances equal to the skin depth $\delta$
of the e.m. fields, as it is usually the case in current Casimir
experiments.

Our analysis of the effect of roughness starts from the remarks
made a long time ago by Pippard in his investigations of the skin
effect \cite{pippard}. Pippard noted that at frequencies in the
GHz region, the presence of microscopic scratches on the surface
of a metal can determine a significant increase of the surface
resistance, by forcing the electric currents to flow along longer,
distorted paths inside the metal. We considered that these remarks
acquire even greater importance if the irregularities of the
surface are smaller than the skin depth $\delta$. If such a
conductor is placed in a tangential alternating electric field,
the induced ohmic currents in the immediate vicinity of the
surface will be unable to follow the rapid distortions of the
surface, with the result that barely any current will be found in
the outer layer of thickness $h$ of the conductor. This is like
water flowing inside a large pipe whose inner surface presents
small bumps. The water contained in these bumps does not take part
in the flow, and remains still. We were thus led to conceive  that
{\it the surface layer of thickness $h$ of a rough conductor does
not behave like the deeper part of the conductor, and in fact,
with regards to tangential electric fields, it rather resembles a
dielectric, having a negligible real conductivity
$\sigma(\omega)$}. A detailed investigation of the response
function of a rough metallic surface, according to the above
picture, is likely to be a rather complicated problem, because
when $h$ is comparable to or less than the electron mean free path
$l$ at room temperature (for gold $l$ is around 20 nm), spatial
dispersion is  expected to be relevant. In this Letter we shall
content ourselves with proposing a simple phenomenological model,
suggested by the previous qualitative considerations. We neglect
the possible effect of spatial dispersion near the metal's
surface, and   we model a (thick)  rough metallic plate as a {\it
plane-parallel two-layer system}, consisting of a  thick slab with
permittivity $\epsilon(\omega)$, describing the inner part of the
conductor, covered by a thin plane-parallel uniform layer of
thickness $h$ with a {\it different} permittivity $\epsilon_{\rm
surf}(\omega)$, modelling its rough surface. For the bulk
permittivity $\epsilon(\omega)$ we take the usual Drude model in
Eq. (\ref{drude}), while for the thin outer layer we take the
plasma model: \be \epsilon_{\rm surf}(\omega)=1-\Omega^2_{{\rm
surf}}/\omega^2\;,\label{plasma}\ee with \be \Omega^2_{\rm surf}=f
\, \Omega^2_{\rm P}\;,\label{fill}\ee where  $f$ is the fraction
of  the plane-parallel  layer of width $h$ really occupied by the
metal. We now explain these choices.  Modelling the rough surface
as a uniform plane-parallel  layer is   reasonable, because in
typical Casimir experiments at room temperature  the wavelengths
$\lambda \simeq a$ and the skin depths $\delta$ of the e.m. fields
are both much larger than  $h$, and therefore the e.m. fields
effectively see the rough surface as flat. The choice of the
plasma-model for $\epsilon_{\rm surf}$ appears as the simplest
one, because the plasma model is what one obtains from the   Drude
model,  after one sets to zero the real part of the conductivity
$\sigma$, in such a way that the surface layer behaves as an
insulator, as we argued it is the case. The chosen value of
$\Omega^2_{\rm surf}$ in Eq. (\ref{fill}) depends on the fact that
the average density $n_{\rm surf}$ of free electrons within the
rough layer is smaller by a factor $f$ than the bulk value $n$.
Upon recalling that the square of the plasma frequency is related
to the density $n$ of free electrons by the well known relation
$\Omega^2=4 \pi n e^2/m$, where $e$ and $m$ denote the electron
and (effective) mass, respectively, one arrives at Eq.
(\ref{fill}).   In principle, for a given rough surface, it should
be possible to compute the values of $h$ and $f$, starting from a
detailed microscopic theory. In practice, we shall regard $h$ and
$f$ as {\it phenomenological} parameters to be determined by
comparison with experimental data.

The two layer structure for a rough metallic plate is what
distinguishes our model from  the models that have appeared in the
recent literature, which all use a single dielectric function
$\epsilon(\omega)$ for the entire plate,  the  various approaches
differing in the choice of $\epsilon(\omega)$. It is clear that
for sufficiently large $h$ our model  is expected to reproduce the
results of the plasma model used in Ref. \cite{decca}, while for
$h \rightarrow 0$ it should resemble the Drude model of
Ref.\cite{sernelius}.
\begin{figure}
\includegraphics{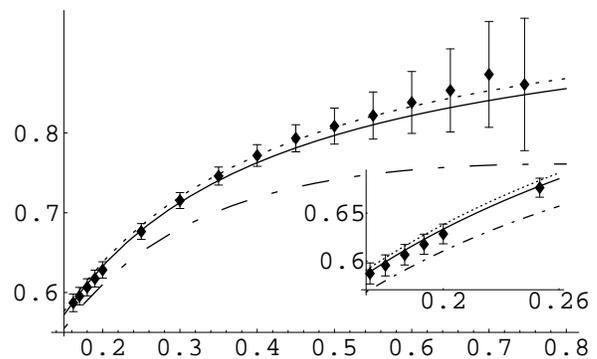}% Here is how to import EPS art
\caption{\label{fig1}  Experimental data for the reduction factor
$\eta$ from Ref. \cite{decca} (diamonds). The inset shows a
magnified view of the picture for   small separations. The solid
and dotted lines are fits using our theory, with $h=11$ nm and
$f=0.9$, for $T=300$ K (solid line) and $T=0$ K (dotted line). The
dot-dashed line is for the Drude model approach at $T=300$ K.
Distances $d$ are in microns. See text for explanations}
\end{figure}
To see exactly how things go, we now compute the thermal Casimir
force between two identical rough plates, using our model. This is
done simply by substituting into Lifshitz formula Eq. (\ref{lifs})
the appropriate reflection coefficients $r_{\alpha}$ for our
two-layer model of a rough plate:  \be
r_{\alpha}=\frac{r_{\alpha}^{(01)}+r_{\alpha}^{(12)}\exp{(-2 h
s^{(1)})}}{1+r_{\alpha}^{(01)}r_{\alpha}^{(12)}\,\exp{(-2 h
s^{(1)}})}\;,\ee where $\epsilon^{(0)}=1$,
$\epsilon^{(1)}=\epsilon_{\rm surf}(\omega)$,
$\epsilon^{(2)}=\epsilon(\omega)$, $s^{(j)}=\sqrt{\epsilon^{(j)}(i
\xi_l) \xi_l^2/c^2+k_{\perp}^2}$, and $r_{\alpha}^{(ij)}$ are the
usual Fresnel reflection coefficients for the interface $ij$. In
what follows we shall plot the Casimir pressure as a function of
the $\it average$ separation $d$ between the plates, that is
related to $a$ as $d=a+2 \, h(1-f)$. Indeed $d$ is a more
meaningful quantity to consider, because it coincides with the
plates separation that is  usually measured in  the experiments.

As a check of our theory, we tried to fit a recent (indirect)
accurate measurement of the Casimir force between two
plane-parallel gold surfaces, using a micromachined oscillator,
reported in Ref. \cite{decca}. The experimental data are shown in
Fig. 1, in terms of the reduction factor $\eta=P/P_{\rm id}$,
where $P_{\rm id}=(\pi^2 \hbar c)/(240 d^4)$ is the Casimir
pressure for two ideal metallic plates, at zero temperature.  In
our computations, we took $\hbar \Omega_P=8.9$ eV and $\hbar
\gamma =.0357$ eV, which are the values used in \cite{decca}. As
the most accurate measurements were performed at rather small
separations, around 200 nm, interband transitions of core
electrons give a sizable contribution to the permittivity, and we
therefore added both to $\epsilon$ and $\epsilon_{\rm surf}$ the
six-oscillator expression $\epsilon_{\rm IB}(\omega)$ accounting
for interband transitions, that was used in Ref.\cite{decca}.
Fig.1 shows two fits with our model for $h=11$ nm and $f=0.9$, for
$T=300$ K (solid-line) and $T=0$ (dotted line). We see that both
curves are consistent with all the measurements in the entire
range of distances, ranging from 162 nm to 746 nm, with errors
well inside the 95 confidence interval reported in Ref.
\cite{decca}. The values for $h$ and $f$ obtained from the fit are
nicely consistent with the degree of roughness of the Au surfaces
reported in Ref. \cite{decca}, which appears  as encouraging in
favor of our model. As we see, the precision of the measurements
is not sufficient to discriminate the $T=0$ and $T=300$ K curves,
and therefore this experiment does not detect the thermal
contribution to the Casimir force.  We remark that the data rule
out the Drude-approach of Ref. \cite{sernelius} (dot-dashed line),
while they are consistent with the plasma-model approach advocated
in Ref. \cite{decca}, and in fact for this experiment our model
gives results that are very close to those obtained by means of
the plasma model.

It is now interesting to consider how the magnitude of the thermal
correction depends on the roughness at larger separations, and in
Fig. 2 we show plots of the reduction factor for different degrees
of roughness.
\begin{figure}
\includegraphics{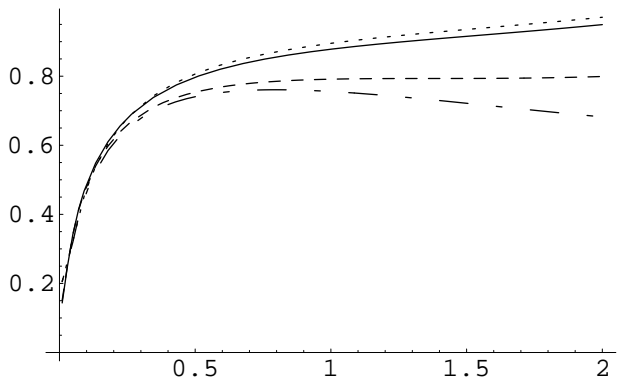}% Here is how to import EPS art
\caption{\label{fig1} Plots of the reduction factor $\eta$ for Au
plates at $T=300$ K, as a function of the {\it average separation}
$d$ (in microns). The solid and dashed curves are for our theory,
with $h=11$ nm and $f=0.9$ (solid curve), $h=2$ nm and $f=0.5$
(dashed curve). The dotted curve is for the plasma-model of Ref.
\cite{decca} and the point-dashed curve for the Drude model of
Ref. \cite{sernelius}.}
\end{figure}
The curves were computed for gold, with $\hbar \Omega_P=8.9$ eV,
and $\hbar \gamma =.0357$ eV, for $T=300$ K.  The  dotted and
dot-dashed line correspond, respectively, to the (generalized)
plasma approach of Ref.\cite{decca} and to the Drude approach of
Ref.\cite{sernelius}. The solid and dashed lines were computed
using our model, with $l=11$ nm, $f=0.9$ (solid) and $l=2$ nm,
$f=.5$ (dashed). Fig. 1 lends itself to several important
comments. First of all we see that, according to our model,
surface roughness does influence strongly the Casimir pressure
also at large separations, contrary to the PFA picture. Second,
neither the plasma model, nor the Drude model are generally valid.
The plasma model may be expected to be valid only for "very" rough
surfaces, having a roughness amplitude $h$ around ten nm, like
those used in the experiment in Ref. \cite{decca}. On the
contrary, the Drude model should be valid for very smooth
surfaces, with a roughness amplitude $h$ well below two nm. For
intermediate values of the roughness amplitude, neither theory
should be adequate and perhaps the simple model proposed in this
Letter provides a better approximation.

In conclusion, we have presented a new phenomenological theory for
the thermal Casimir effect between two rough metallic plates.
According to this theory, surface roughness influence strongly the
thermal correction to the Casimir force at all separations $a$.
The theory is expected to be valid for surfaces with a roughness
having a characteristic spatial scale $h$ below the skin depth
$\delta$, which represents the standard experimental situation.
For smoother surfaces, having slow profile variations on the scale
of the skin depth, the usual approximations based on the PFA are
expected to be correct, and then roughness should only give a
small correction of order $(h/a)^2$.  We do not know yet if our
two-layer model is free from the thermodynamical inconsistencies
at low temperature that plague the Drude model \cite{bezerra}, and
we plan to investigate this important problem in a future work.
Another important remark is that the model presented in this
Letter can be used to obtain predictions for other proximity
effects, originating from thermal e.m. fluctuations, like for
example the power $S$ radiative heat transfer between two metallic
plates at different temperatures, separated by an empty gap.
Investigating this problem would be very interesting, because that
would provide us with another check of the correctness of our
model for a rough metallic surface. This subject also will be
addressed in future work.

\end{document}